\documentclass[pra,aps,a4paper, twocolumn,repreprint, superscriptaddress, longbibliography]{revtex4-2}

\usepackage{amssymb}
\usepackage{amsmath}
\usepackage{amsfonts}
\usepackage{graphicx}
\usepackage{color, soul}
\usepackage{natbib}
\usepackage{hyperref}
\usepackage{adjustbox}
\usepackage{tabularx}
\usepackage{breqn}
\usepackage{multirow}
\usepackage{lineno}

\begin{document}

     \title{Low-Temperature Heat Transport under Phonon Confinement in Nanostructures}

	\author{M. Sidorova}
    \email{Mariia.Sidorova@dlr.de}
    \affiliation{Nanyang Technological University$,$School of Physical and Mathematical Science $,$ 21 Nanyang Link $,$ 637371 Singapore}
	\affiliation{Humboldt-Universität zu Berlin$,$ Department of Physics$,$ Newtonstr. 15$,$ 12489 Berlin$,$ Germany}
	\affiliation{German Aerospace Center (DLR)$,$ Institute of Optical Sensor Systems$,$ Rutherfordstr. 2$,$ 12489 Berlin$,$ Germany}
	
	\author{A.D. Semenov}
	\affiliation{German Aerospace Center (DLR)$,$ Institute of Optical Sensor Systems$,$ Rutherfordstr. 2$,$ 12489 Berlin$,$ Germany}

    \author{A. Zaccone}
	\affiliation{Department of Physics “A. Pontremoli”$,$ University of Milan$,$ via Celoria 16$,$ 20133 Milan$,$ Italy}
	\affiliation{Institute for Theoretical Physics$,$ University of Göttingen$,$ Friedrich Hund Platz 1$,$ 37077 Göttingen$,$ Germany}

	\author{I. Charaev}
	\affiliation{Physics Institute$,$ University of Zürich$,$ Winterthurerstrasse 190$,$ 8057 Zürich$,$ Switzerland}

    \author{M.  Gonzalez}
	\affiliation{Physics Institute$,$ University of Zürich$,$ Winterthurerstrasse 190$,$ 8057 Zürich$,$ Switzerland}

	\author{A. Schilling}
	\affiliation{Physics Institute$,$ University of Zürich$,$ Winterthurerstrasse 190$,$ 8057 Zürich$,$ Switzerland}

    \author{S. Gyger}
	\affiliation{Department of Applied Physics$,$ KTH Royal Institute of Technology$,$ SE-106 91 Stockholm$,$ Sweden}
    \affiliation{Ginzton Laboratory$,$ Stanford University$,$ 348 Via Pueblo Mall$,$ Stanford$,$ California 94305$,$ USA}
	
	\author{S. Steinhauer}
	\affiliation{Department of Applied Physics$,$ KTH Royal Institute of Technology$,$ SE-106 91 Stockholm$,$ Sweden}

	\begin{abstract}
	
	Heat transport in bulk materials is well described using the Debye theory of 3D vibrational modes (phonons) and the acoustic match model. However, in cryogenic nanodevices, phonon wavelengths exceed device dimensions, leading to confinement effects that standard models fail to address. With the growing application of low-temperature devices in communication, sensing, and quantum technologies, there is an urgent need for models that accurately describe heat transport under confinement. We introduce a computational approach to obtain phonon heat capacity and heat transport rates between solids in various confined geometries, that can be easily integrated into, e.g., the standard two-temperature model.
 Confinement significantly reduces heat capacity and may slow down heat transport. We have validated our model in experiments on strongly disordered NbTiN superconducting nanostructures, widely used in highly efficient single-photon detectors, and we argue that confinement is due to their polycrystalline granular structure. These findings point to potential advances in cryogenic device performance through tailored material and microstructure engineering.
	\end{abstract} 

\date{\today}
\maketitle

\section{Introduction}
\label{sec: Introduction}

Low-temperature devices, including superconducting and hybrid quantum sensors for particles \cite{BassDoser2024} and photons\cite{esmaeil2021superconducting}, circuits\cite{devoret2013superconducting}, and electronics\cite{braginski2019superconductor} have become integral to quantum technology, detection, sensing, and metrology. For instance, superconducting nanostrip single-photo detectors (SNSPDs, commonly known as nanowire detectors) cascaded into kilo-pixel arrays enable quantum imaging \cite{oripov2023superconducting}, while advanced superconducting digital electronics \cite{Tolpygo2016} with sophisticated layering and miniaturization are expected to outperform CMOS technology in high-performance computing. However, as device complexity grows, so does excess heat dissipation coupled with cryogenic limitations in heat removal, driving these devices beyond their optimal operation. While heat transport in crystalline bulk materials is well understood through the Debye theory of 3D vibrational modes (phonons) \cite{little1959transport}, in cryogenic nanodevices, phonon wavelengths approach device dimensions causing confinement, which standard heat transport models fail to address.

Confinement imposed by device geometry restricts vibrational modes, altering the thermal properties of solids and affecting heat transport at a large scale. Experimental works have demonstrated various confinement effects, such as changes in the frequency scaling of vibrational density of states in a few nanometer-thick films of amorphous ice at 120~K \cite{yu2022omega}; reduced thermal conductivity in sub-20~nm (Ge–Si core-shell) nanowires between 100 and 300~K \cite{wingert2011thermal}; enhanced specific heat in 2~nm lead grains below 10~K \cite{novotny1972effect}. Advanced thermal imaging techniques, like SQUID-on-tip thermometry, have spatially resolved heat dissipation at individual defects in 2D graphene \cite{halbertal2016nanoscale, halbertal2017imaging}.

Sparse theoretical efforts addressed the problem. Early studies\cite{baltes1973specific, lautehschlager1975improved} derived phonon heat capacity by counting vibrational modes in spherical grains, showing that surface quality (suspended versus clamped) significantly impacts heat capacity through the presence or absence of surface modes. Later works\cite{prasher1999non, mcnamara2010quantum} expanded this to nanostructures of varying sizes and shapes. The vibrational density of states in thin films was derived in Ref. \cite{yu2022omega} based on the confinement model from Ref. \cite{phillips2021universal}.
The same model \cite{phillips2021universal} also explains diverse phenomena such as shear elasticity in confined liquids \cite{Trachenko2020}, suppression of THz modes and reduced dielectric permittivity in nanoconfined media such as water and ferroelectric films \cite{Park2024, ZacconePRB2024}, as well as reduced heat capacity in thin films \cite{Zaccone2024} 
(see a comment \footnote{In Ref.\cite{Zaccone2024}, the analysis initially showed good agreement between experimental heat capacity data from our lab and the analytical $T^4 d$ law for thin films. However, later we revealed discrepancies, as the heat capacities were indirectly derived from electron cooling times using the two-temperature model. In particular, the original analysis overlooked the confinement effect on electron cooling. By incorporating this effect, we achieved quantitative consistency with experimental data, assuming tree-dimensional confinement in grains as detailed in the present paper.}).
More recent works described heat removal for metallic grains embedded in dielectric matrices \cite{bezuglyj2024phonon} and heat dissipation in moderately disordered mesoscopic systems\cite{kong2018resonant, tikhonov2019asymmetry, leumer2024going}, showing enhanced electron-phonon cooling due to scattering at individual defects and a shift in Joule heating and electron cooling profiles relative to the defect center. Despite these efforts, a comprehensive model for heat transport under confinement across diverse geometries and materials is still lacking.

In this work, we derive explicit expressions for phonon heat capacity and heat transport rates between solids for various confined geometries. We validate our approach using experimental data from strongly disordered NbTiN superconducting nanostructures, widely used in highly efficient single-photon detectors \cite{esmaeil2021superconducting}. We also provide physical insights on how confinement affects cryogenic nanodevice performance, suggesting strategies for optimization.

\begin{figure*}[t!]
	\centerline{\includegraphics[width=0.95\textwidth]{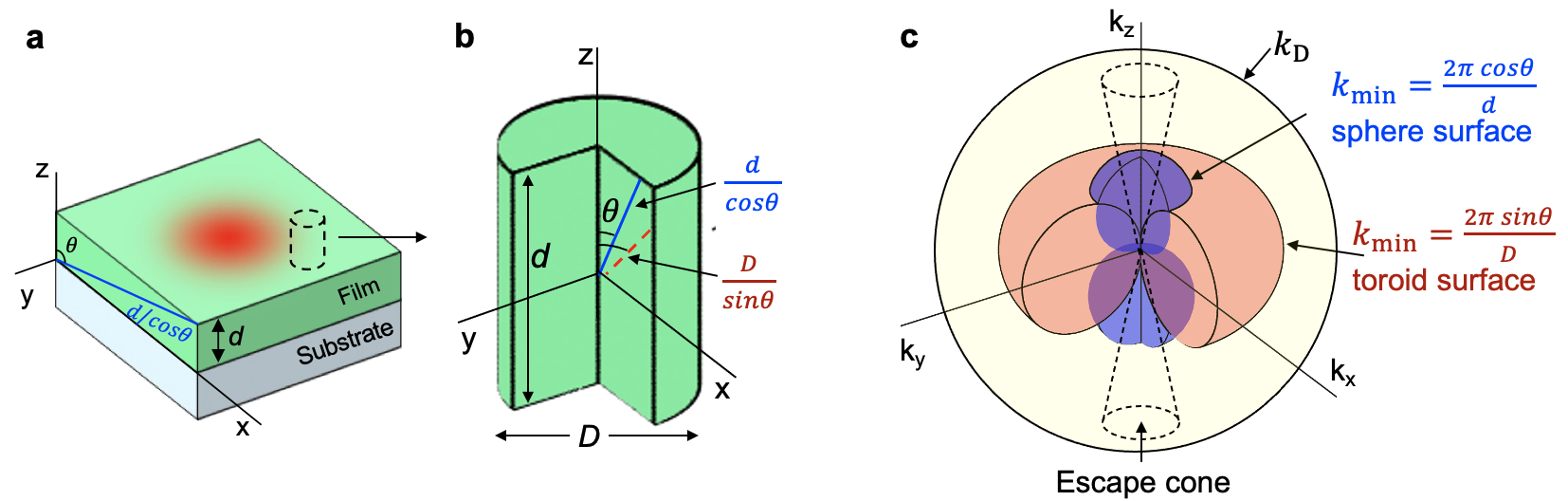}}
	\caption{Model of heat transport under phonon confinement. (a) Real-space schematic of film with thickness \textit{d}. Film dimensions in the \textit{xy} plane are much larger than in the \textit{z} direction (b) Real-space schematic of a cylindrical sample (or nanowire) with height $d$ in the \textit{z} direction and diameter \textit{D} in the \textit{xy} plane. For both, (a) and (b), the angle $\Theta$ is measured from the \textit{z} axis, and rotational symmetry is maintained in the \textit{xy} plane. (c) The wavevector \textit{k}-space corresponds to the confined geometries shown in (a) and (b). Confinement imposed by film planes (along the \textit{z} direction) and by cylinder walls (in the \textit{xy} plane) prohibits phonon states within blue spheres and a red toroid, respectively, inside the Debye (yellow) sphere. The escape cone is defined by the total internal reflection of phonons at the interface with the substrate according to the acoustic match model \cite{kaplan1979acoustic}.}
	\label{fig:real_k_space}
\end{figure*}

\section{Model of heat transport under phonon confinement}
\label{sec: Model}
Below we address confinement in thin films and cylinders, dragging an idea from \cite{phillips2021universal}. We derive expressions for the phonon heat capacity and the escape time of phonons from the sample to the underlying substrate.

\subsection{Phonon heat capacity}

For a bulk 3D crystal, the vibrational (phonon) heat capacity per unit volume and per one polarization is given by
\begin{multline}\label{eq:c_3D}
    c_{ph}^{{\langle3D\rangle}}(T) = \frac{\partial U}{\partial T} = \int_{0}^{\omega_D} \frac{\partial}{\partial T}\left(\frac{g(\omega) \hbar\omega}{e^{\hbar \omega/k_B T}-1} \right) \, \text{d}\omega\\
    = \frac{1}{2\pi^2} k_B \left(\frac{k_B T}{\hbar u}\right)^3 \int_{0}^{x_D} \, \frac{x^4e^x}{(e^x-1)^2} \, \text{d}x,
\end{multline}
where $g(\omega)=\omega^2/(2\pi^2 u^3)$ is the Debye density of states (DOS, per unit volume), $k_B$ and $\hbar$ are the Boltzmann and reduced Planck constants, $\omega_D=(6\pi^2 u^3/a_0^3)^{1/3}$ is the Debye angular frequency, $u$ and $a_0$ denote the sound velocity and the lattice constant, $x=\hbar \omega / k_B T$. Given the linear dispersion $\omega=u k$ ($k$ is the wavevector), valid at temperatures much lower than the Debye temperature, and a low-temperature approximation $x_D\to \infty$, the integral in Eq.~(\ref{eq:c_3D}) simplifies to $4\pi^4/15$. This leads to the well-known Debye result: $c_{ph}^{<3D>}=(2\pi^2/15)k_B (k_B T/\hbar u)^3$.

\subsubsection{$c_{ph}$, confinement in thin film}
For thin films with geometry depicted in Fig.~\ref{fig:real_k_space}a, confinement is imposed by film plains. It restricts phonon states with wavelengths larger than $d/\cos\Theta$, where $d$ is the film thickness, and $\Theta$ is the angle measured from the confinement \textit{z}-axis.

In spherical coordinates of the wavevector $k$-space, illustrated in Fig.~\ref{fig:real_k_space}c, the wavevectors of restricted phonon states $k_{min}=2\pi \cos\Theta/d$ define the surface of blue spheres within the Debye sphere. Consequently, this modifies heat capacity that can be described by introducing an angle-dependent integral limit and integrating over the polar angle $\Theta$:
\begin{multline}\label{eq:c_film}
    c_{ph}^{\langle2D\rangle}(T, d) = \frac{1}{2\pi^2} k_B \left(\frac{k_B T}{\hbar u}\right)^3 \\
    \times \int_0^{\pi/2} \sin\Theta \,\int_{x^*(\Theta)}^{x_D} \, \frac{x^4e^x}{(e^x-1)^2} \, \text{d}x \, \text{d}\Theta ,
\end{multline}
where, $x^*(\Theta)=2\pi \cos\Theta \, \hbar u  / (d\,k_B \,T) $.

\begin{figure}
    \centerline{\includegraphics[width=0.45\textwidth]{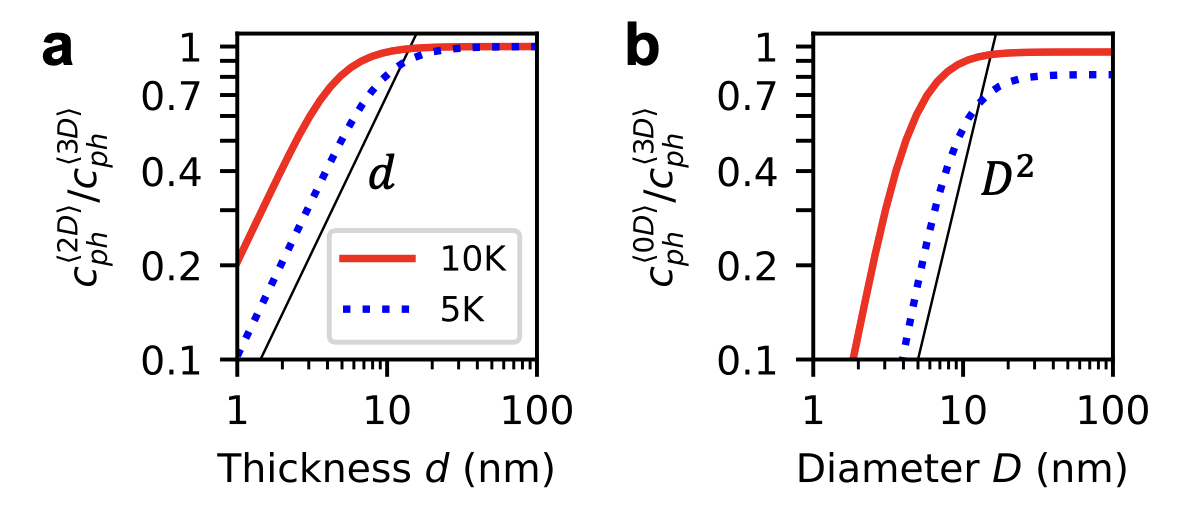}}
    \caption{Reduced phonon heat capacities normalized to the Debye value for confinement imposed by (a) thin film, Eq.~(\ref{eq:c_film}), and (b) cylinder with a fixed height of 10~nm, Eq.~(\ref{eq:c_cylinder}). The legend in panel (a) applies to both panels, indicating fixed temperatures. Thin black lines show the expected scaling for the two confined geometries. Material parameters used correspond to those of NbTiN.}
    \label{fig:cphs_model}
\end{figure}

Alternatively, one can account for the confinement by modifying the phonon DOS inside a sphere with radius $2\pi /d$ that has been done in previous studies\cite{phillips2021universal, travaglino2022analytical, yu2022omega}, which results in a different expression for the heat capacity Eq.~(\ref{eq:c_film_alter}) in Appendix. In the limit of strong confinement ($x^*(0) \to x_D$ in Eq.~(\ref{eq:c_film_alter})), this alternative approach clearly shows that heat capacity reduces upon decreasing the film thickness and temperature as $c_{ph}^{<2D>} \propto d \, T^4$. In the bulk limit, where $x(\Theta)^* \to 0$ in Eq.~(\ref{eq:c_film}), one retrieves the Debye result. These two limiting cases are seen in Fig.~\ref{fig:cphs_model}a.

\subsubsection{$c_{ph}$, confinement in cylinder}
A cylindrical confined geometry with a height $d$ and diameter $D$ is illustrated in Fig.~\ref{fig:real_k_space}b. This geometry could be relevant to nanowires (cylinders with infinite height, a confined 1D structure) as well as to films with granular morphology (densely packed confined 0D grains). Cylinder bases restrict phonon states with wavelengths larger than $d/\cos\Theta$ for angles $\Theta \leq\tan^{-1}(D/d)$; we note that for $D=d$, $\Theta=45^{\circ}$. In addition, cylinder walls restrict phonon states with wavelengths larger than $D/\sin\Theta$ for larger angles $\Theta \geq \tan^{-1}(D/d)$. 

In the \textit{k}-space, the wavevectors of phonon states restricted by cylinder bases ($k_{min}=2\pi \cos\Theta/d$) and walls ($k_{min}=2\pi \sin\Theta/D$) with corresponding conditions on angles, define the surface of blue spheres and red toroid, respectively (see Fig.~\ref{fig:real_k_space}c).  These restrictions modify the phonon heat capacity as follows:
\begin{multline}\label{eq:c_cylinder}
    c_{ph}^{\langle0D\rangle}(T, d, D) = \frac{1}{2\pi^2} k_B \left(\frac{k_B T}{\hbar u}\right)^3 \\\times \biggl[
    \int_{0}^{\tan^{-1}(D/d)} \sin\Theta \,\int_{x^*(\Theta)}^{x_D} \, \frac{x^4e^x}{(e^x-1)^2} \, \text{d}x \, \text{d}\Theta \\
    + \int_{\tan^{-1}(D/d)}^{\pi/2} \sin\Theta \int_{x^{**}(\Theta)}^{x_D} \, \frac{x^4e^x}{(e^x-1)^2} \, \text{d}x \,\text{d}\Theta  \biggr] .
\end{multline}
Here, the lower limit $x^*(\Theta)$ is caused by the confinement along the $z$ direction, and $x^{**}(\Theta)=2\pi \sin\Theta \, \hbar u  / (D\,k_B\,T) $ in the $xy$-plane. 

As the cylinder diameter becomes much larger than its height, $D\gg d$, the integral limits in Eq.~(\ref{eq:c_cylinder}) $\tan^{-1}(D/d)\to \pi/2$, and the second term diminishes, retrieving the result for confinement in films. In the opposite limit of strong confinement by the cylinder walls, $D\gg d$, $\tan^{-1}(D/d)\to 0$, and the first term diminishes, while the second integral in the second term reduces fast as $x^{**}(\Theta)\to x_D$. These cases are shown in Fig.~\ref{fig:cphs_model}b.

\subsection{Phonon escape time}

Heat transport between two dissimilar bulk solids was described in \cite{little1959transport}.
The heat flow per unit area from the film (not indexed) to the substrate (index 0) via phonons (one polarization) is given by
\begin{multline}\label{eq:dQdt}
    \frac{\text{d}Q}{\text{d}t} = \frac{u}{2} \iint \frac{g(\omega) \hbar\omega}{e^{\hbar \omega/k_B T}-1} \, \alpha(\Theta) \sin\Theta\cos\Theta \text{d}\omega \, \text{d}\Theta  \\
        - \frac{u_0}{2} \iint  \frac{g_0(\omega) \hbar\omega}{e^{\hbar \omega/k_B T_0}-1}  \, \alpha(\Theta_0) \sin\Theta_0\cos\Theta_0 \text{d}\omega \, \text{d}\Theta_0\\
        =\frac{u}{2} \left[ \mathcal{I}(T)-\mathcal{I}(T_0) \right].
\end{multline}
Here, $\alpha(\Theta)$ is the phonon transmission coefficient through the interface, given by the acoustic match model \cite{kaplan1979acoustic}, which depends on the phonon angle of incidence $\Theta$  and acoustic impedances of the two media. For many solid/solid interfaces, $\alpha(\Theta)$ is constant at small angles and drops to zero at the critical angle of total internal phonon reflection, depicted by the escape cone in Fig.~\ref{fig:real_k_space}c. Phonon states with wavevectors beyond this cone are reflected at the interface and can only escape the film through mechanisms such as inelastic scattering (e.g., absorption and re-emission by electrons) or mode conversion at the interface. 

In Eq.~(\ref{eq:dQdt}), since the integrals over frequency are angle-independent, the double integrals $\mathcal{I}(T)$ can be separated. They yield products of the angle-averaged transmission coefficient $\langle\alpha\rangle=2\int_0^{\pi/2} \alpha(\Theta) \sin\Theta\cos\Theta \,\text{d}\Theta$ and the phonon energy $U = \frac{1}{2\pi^2} k_B T \left(\frac{k_B T}{\hbar u}\right)^3 \int_{0}^{x_D} \, \frac{x^3}{(e^x-1)} \, \text{d}x$. With the low-temperatures approximation $x_D \to \infty$, the heat flow becomes $\text{d}Q/\text{d}t = \frac{\pi^2}{60} \langle\alpha\rangle \frac{k_B^4 }{\hbar^3 u^2} (T^4-T_0^4)$.

Denoting $\Delta T=(T-T_0)$ and assuming small temperature differences $\Delta T\ll (T, T_0)$, the heat flow can be linearized as follows
\begin{equation}\label{eq:dQdt_lin}
    \frac{\text{d}Q}{\text{d}t} = \frac{u}{2} \frac{\text{d}\left[\mathcal{I}(T)\right]}{\text{d}T}\Delta T.
\end{equation}
On the other hand, using $\tau$-approximation for small deviations from equilibrium $\text{d}x/\text{d}t=-x/\tau$ (see e.g. \cite{bezuglyi1997kinetics} or "correlations of fluctuations in time" in \cite{landau2013statistical}), one can express the heat flow via the rate of energy escape through the interface ($1/\tau_{es}$) as $\text{d}Q/\text{d}t = (2d \, c_{ph}/\tau_{es}) \Delta T$. Here, factor 2 appears because the heat flows through only one boundary. By plugging this into Eq.~(\ref{eq:dQdt_lin}), the linearized phonon escape time is given by
\begin{equation}\label{eq:tau_es}
    \tau_{es}^{\langle3D\rangle} = \frac{4d}{u} \frac{c_{ph}(T)}{\text{d}\left[\mathcal{I}(T)\right]/\text{d}T} = \frac{4d}{\langle\alpha\rangle u},
\end{equation}
where for bulk material the derivative $\text{d}\left[\mathcal{I}(T)\right]/\text{d}T=\langle\alpha\rangle c_{ph}(T)$.

\subsubsection{$\tau_{es}$, confinement in thin films}

Using the formalism described above, we derive the phonon escape time under confinement in one medium imposed by a thin film, the geometry is shown in Fig.~\ref{fig:real_k_space}a. Only phonon states with wavelengths smaller than  $d/\cos\Theta$ (wavevectors $k_{min}=2\pi \cos \Theta/d$) are present in the film and participate in heat transport, carrying energy out of the film. Their density of states is the Debye DOS $g(\omega)$. Accordingly, this modifies the low-frequency limits of double integrals in Eq.~(\ref{eq:dQdt}).  

As evident from Fig.~\ref{fig:real_k_space}c, spheres that prohibit phonon states with $k_{min}$ strongly overlap with the phonon escape cone. Confinement in film reduces the number of phonon states, which could directly escape to the substrate, slowing down heat transport compared to the bulk limit.

The modified phonon escape time (see Appendix~\ref{app: tau_es}), is given by 
\begin{equation}\label{eq:tau_es_2d}
    \tau_{es}^{\langle2D\rangle} = \frac{4d}{\langle\alpha\rangle u} \, \frac{c_{ph}^{\langle2D\rangle}(T, d)}{c_{ph}^{\langle3D\rangle}(T) \left[1-\mu(T, d)\right]},
\end{equation}
where  $\mu(T, d) = \frac{15}{4\pi^4} \int_0^{\pi/2} \alpha(\Theta) \sin\Theta\cos\Theta \text{d}\Theta \cdot \int_{0}^{x^*(\Theta)} \frac{x^4e^x}{(e^x-1)^2}\, \text{d}x$, and $x^*(\Theta)=2\pi \cos\Theta /d \,\, \hbar u  / k_B T $. 

\begin{figure}
    \centerline{\includegraphics[width=0.45\textwidth]{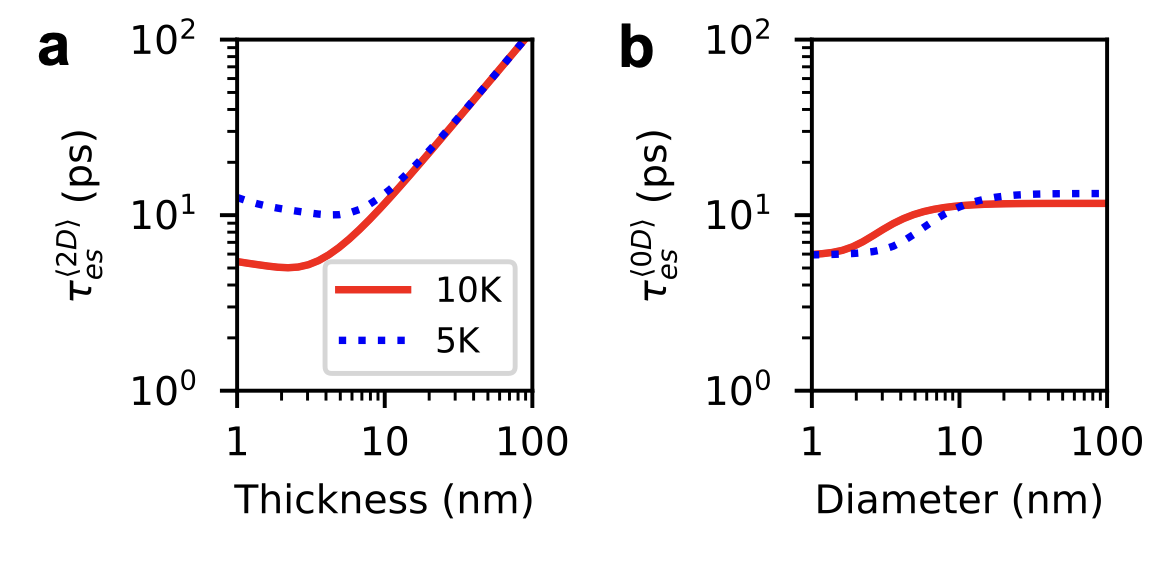}}
    \caption{Modified phonon escape times for confinement imposed by (a) thin film, Eq.~(\ref{eq:tau_es_2d}), and (b) cylinder with a fixed height of 10~nm, Eq.~(\ref{eq:tau_es_1d}). The legend in panel (a) applies to both panels, indicating fixed temperatures. Material parameters used correspond to those of NbTiN/SiO$_2$.}
    \label{fig:taus_model}
\end{figure}

In the bulk limit, $k_{min} \to 0$ for any $\Theta$, the blue sphere volume $\to 0$ in Fig.~\ref{fig:real_k_space}c, retrieving the 3D result. In the limit of strong confinement, there are fewer allowed phonon states participating in heat transport which dramatically slows it down. These two limits are seen in Fig.~\ref{fig:taus_model}a.

\subsubsection{$\tau_{es}$, confinement in cylinders}

Following a similar approach, we derive the phonon escape time for a cylindrical sample, Fig.~\ref{fig:real_k_space}b, where phonons escape through one of the cylinder bases to the substrate. For angles $<\Theta^*= \tan^{-1}(D/d)$, confinement is imposed by cylinder bases (as for the film), and for angles $>\Theta^*$ by cylinder walls. For simplicity, we ignore phonon scattering at the wall boundaries (see discussion).

The impact of such additional confinement by cylinder walls depends on the ratio between $\Theta^*$ and the critical angle of total internal reflection $\Theta_C$ ($\Theta_C$ defines the escape cone in Fig.~\ref{fig:real_k_space}b). 
For  $\Theta_C<\Theta^*$ the walls prohibit phonon states only beyond the escape cone (otherwise these states could not directly escape the film but via inelastic scattering or mode conversion), speeding up phonon escaping. In the opposite case $\Theta_C>\Theta^*$, cylinder walls do not restrict heat transport and therefore do not affect the escape time. Most likely, however, cylinder diameters do not exceed the film thickness, thus $\Theta^* \leq 45^\circ$, while acoustic impedances of typical nanodevices/substrates result in a wide range of $\Theta_C \geq 30-70^\circ$.

The phonon escape time for a cylindrical sample is given by:
\begin{equation}\label{eq:tau_es_1d}
    \tau_{es}^{\langle0D\rangle} = \frac{4d}{\langle\alpha\rangle u} \, \frac{c_{ph}^{\langle0D\rangle}(T, d, D)}{c_{ph}^{\langle3D\rangle}(T) \left[1-\mu(T, d, D)\right]},
\end{equation} 
where $\mu(T, d, D) = \frac{15}{4\pi^4} \int_0^{\pi/2} \alpha(\Theta) \sin\Theta\cos\Theta \text{d}\Theta 
    \cdot \left[ \int_{0}^{x^*(\Theta)} H_\Theta \frac{x^4e^x\, \text{d}x}{(e^x-1)^2} + \int_{0}^{x^{**}(\Theta)} (1-H_\Theta) \frac{x^4e^x\, \text{d}x}{(e^x-1)^2} \right]$, the upper limits $x^*(\Theta)$ and $x^{**}(\Theta)$ were defined previously. Here $H_\Theta$ is a step function, $H_\Theta=1$ for $\Theta\leq \tan^{-1}(D/d)$ and $H_\Theta=0$ for $\Theta > \tan^{-1}(D/d)$. 

In the limit $D\gg d$, the volume occupied by toroid in \textit{k}-space (see Fig.~\ref{fig:real_k_space}c) diminishes, and the 2D result is retrieved. In the limit of strong confinement by cylinder walls $D\ll d$, there are fewer phonon states outside the phonon escape cone which speeds up their escaping. These two limiting cases are shown in Fig.~\ref{fig:taus_model}.

\section{Results: Comparison with Experimental Data}

\begin{figure}[t!]
	\centerline{\includegraphics[width=0.45\textwidth]{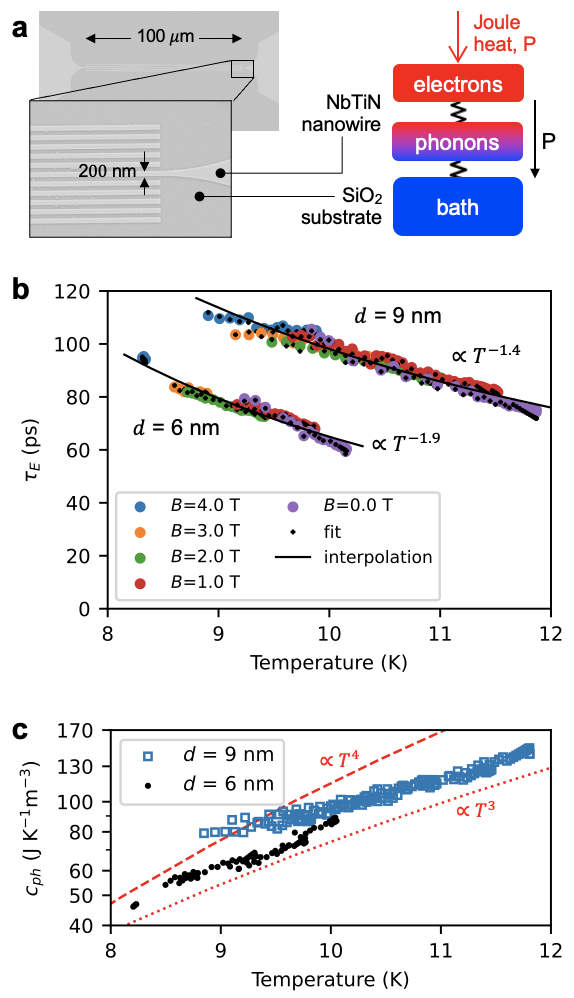}}
	\caption{ Experimental analysis of heat transport in NbTiN devices on SiO$_2$ substrate at low temperatures. (a) Scanning electron microscope images of NbTiN nanostrips, along with schematic heat transport: from electrons to phonons within the nanostrip and subsequently to the thermal bath (substrate). (b) Steady-state electron energy relaxation time was obtained for the NbTiN devices with thicknesses $d=6$ and 9~nm indicated. Experimental data (colored circles) are overlaid with best fits (black dots) obtained with Eq.~(\ref{eq:ss_time}). (c) Phonon heat capacities (markers), computed for confined cylindrical geometry, are reduced to 0.1 times the Debye prediction for bulk material (details are in the main text). The dotted and dashed lines scale as Debye $T^3$ law and analytic $T^4$ law for confined thin films, respectively.}
	\label{fig:exp1}
\end{figure}

We applied our model to describe heat transport in strongly disordered NbTiN superconducting devices with the geometry depicted in Fig.~\ref{fig:exp1}a.

Heat transport was studied in a \textit{steady state} using the well-established method of self-heating normal domain (see details in Appendix~\ref{app_SHNM}). To expand the experimental temperature range, we adapted this method to non-zero magnetic fields. Magnetic field reduces the superconducting transition temperature, enabling the heat transport to be studied at lower temperatures. This method involves heating electrons via Joule heat (denoted as \textit{P} in Fig.~\ref{fig:exp1}a), generated by a direct current. The heat flows from electrons to phonons coupled via electron-phonon interaction and then to the thermal bath via phonon escaping to the substrate. We experimentally obtained the overall time required for electron energy to relax from the nanostrip to the substrate $\tau_E$. These times, plotted in Fig.~\ref{fig:exp1}b with colored circles, increase with the nanostrip thickness and with lowering the temperature.

The steady-state electron energy relaxation time $\tau_E$ is given by the following equation derived from the two-temperature model (for more details see the elaborated model in Ref.\cite{SemenovAnalytical2024}):
\begin{equation}\label{eq:ss_time}
    \tau_E(T_e)=\tau_{EP}(T_e) + \gamma (c_e(T_e)/c_{ph}(T_{ph}))\tau_{es}
\end{equation}
Here, $T_e$ and $T_{ph}$ denote the effective temperatures of electrons and phonons, respectively, $c_e$ and $c_{ph}$ denote their heat capacities, and $\gamma$ is about one (see Appendix). The electron-phonon interaction time $\tau_{EP}$ in thin disordered NbTiN films scales as $T^{-n}$ with $n=3.4-3.5$, ranging from 5 to 1~ps over the temperature range of 8-12~K \cite{sidorova2021magnetoconductance}.

To fit the experimental $\tau_E(T)$ data, we used known NbTiN material parameters\cite{sidorova2021magnetoconductance}, such as $c_e$ and $\tau_{EP}$, while the phonon heat capacity $c_{ph}$ and escape time $\tau_{es}$ were computed with our model accounting for three phonon polarizations. Given the level of disorder in granular NbTiN films \cite{sidorova2021magnetoconductance}, which likely impacts the shear modulus \cite{Zaccone2011, krief2024spatial}, we treated the transverse sound velocity as a fitting parameter. Optimal agreement with experimental data was only achieved by assuming confinement in all three directions, i.e., confined cylindrical geometry. Using $c_{ph}$ and $\tau_{es}$ computed for 3D bulk phonons, as well as for phonons confined in a film, resulted in relaxation time values that were underestimated, by at least a factor of four, compared to the experimental values. The cylindrical geometry is likely due to the polycrystalline granular morphology of NbTiN films, where crystalline grains (nanometric cylinders) are embedded in an amorphous environment \cite{Zichi:19}. A similar dense columnar granular morphology has been reported in thicker NbTiN films \cite{KARIMI201614}.

The best-fit relaxation times are plotted in Fig.~\ref{fig:exp1}b, with black dots. These were obtained with sound velocities reduced to $0.94\pm 4.4\%$ and $0.95\pm 7.1\%$ times the bulk transverse sound velocity of $4700$~m/s \cite{arockiasamy2016ductility} for the 6 and 9~nm-thick strips, respectively. The reduced velocities are consistent with the structural difference between a bulk monocrystal and an aggregate of crystalline grains at the nanoscale, since structural disorder always reduces the shear modulus (compared to the corresponding perfect crystal) due to nonaffine displacements \cite{Zaccone2011}. The longitudinal velocity was fixed to 8400~m/s, the same as in the bulk material\cite{arockiasamy2016ductility}, consistent with the well-known fact that the compression modulus is less affected by disorder and nonaffinity \cite{Zaccone_book}. We also varied the cylinder diameter (\textit{D}), while fixing its height to equal the film thickness. The best-fit diameter for both devices was 1.72~nm, containing at least four elementary cells along the confined direction. Strictly, this low number of elementary cells requires replacing integration with summation over allowed modes. However, for the sake of the self-consistency of our model, we keep using integral representation for all geometries. Moreover, given the unknown cross-sectional shapes of columnar grains, the distribution of their sizes, and the packing density, our best-fit value should not be considered as a physical grain diameter.

Using these parameters, we computed phonon heat capacities with Eq.~(\ref{eq:c_cylinder}), which are plotted in Fig.~\ref{fig:exp1}c (markers). They are significantly reduced, $\approx0.1$ times the Debye prediction for 3D bulk material, and exhibit a clear dependence on film thickness. For reference, Fig.~\ref{fig:exp1}c shows the Debye $T^3$ scaling (dotted line) and the analytical $T^4$ law (dashed line, Eq.~(\ref{eq:c_film_alter})) for confined thin films \cite{Zaccone2024}, leaving the discussion of the temperature dependence of heat capacity outside the scope of this work. We also computed phonon escape times with Eq.~(\ref{eq:tau_es_1d}), obtaining values of $\tau_{es}=4.2\pm2.1\%$~ps and $6.1\pm4.2\%$~ps for the 6~ and 9~nm-thick devices, respectively. According to Eq.~(\ref{eq:ss_time}), these escape times, along with large $c_e/c_{ph}$ ratios (due to strongly reduced $c_{ph}$), lead to large energy relaxation times $\tau_E$. The computed $\tau_{es}$ scale linearly with thickness as $\approx0.7 d$ and are slightly smaller compared to what is expected in the bulk limit, $\approx1.1 d$ (the film/substrate interface transmission
coefficient is $\approx0.7$). The difference is due to the confinement effect imposed by cylinder walls.


\section{Discussion}

Below we evaluate the implications of our findings on phonon mode confinement and discuss how these insights can advance the understanding and performance of nanodevices at low temperatures.

\subsection*{Validation of the Heat Transport Model}
Our heat transport model describes the effect of confinement imposed by the geometry of nanostructures on phonon modes. The reduction in allowed phonon modes not only (i) reduces phonon heat capacities and (ii) modifies phonon escape time but also (iii) significantly slows down electron energy relaxation (overall electron cooling), as demonstrated through our application of the two-temperature model via Eq.~(\ref{eq:ss_time}).

Experiments on NbTiN devices support our assumption that phonon modes are confined within polycrystalline columnar grains resembling cylinders. The overall cylinder sizes of a few nanometers are comparable to phonon mean-free paths. In this case, as shown in \cite{bezuglyj2024phonon}, the rate of heat removal is not governed by reflections at the grain boundaries (cylinder walls), but rather by the electron-phonon interactions within grains.

Our model is based on the assumption that low-frequency (long-wavelength) phonons do not contribute to heat transport due to confinement effects. This assumption is unequivocally valid when the confined medium interfaces with a vacuum. However, when it interfaces with another solid, such as a substrate, the vibrational wave can extend into the substrate despite acoustic mismatch at the interface. Consequently, the criteria for determining the low-frequency cutoff become less clear. It is even less defined when phonon modes are confined within crystalline granules of polycrystalline films, where the grain boundaries are often thinner than the phonon wavelength. Addressing this issue comprehensively would greatly complicate our heat transport model, shifting the focus from its convenient practical application. However, we speculate that our simplified model remains valid because the grain boundaries are likely oxidized, and the grain sizes serve as an adjustable parameter.

Effects shown in moderately disordered 2D graphene \cite{kong2018resonant, tikhonov2019asymmetry, leumer2024going}, such as impurity-enhanced electron-phonon scattering and temperature profile asymmetries near defects, should also be present in our NbTiN devices, but they do not impact our results. Our devices are strongly disordered, where impurities are densely and uniformly distributed over a distance of a few nanometers.
In our experiments with self-heating normal domains with lengths of tens of micrometers, measured heat transport is dominated by the central domain region, rather than by non-trivial scattering effects around narrow domain walls.

\subsection*{Implications for Nanodevices}
 
The effect of confined phonon modes can significantly enhance the performance of SNSPDs. As confinement increases, $c_{ph}$ decreases, resulting in higher sensitivity to low-energy photons. This sensitivity enhancement is due to the increased fraction of photon energy transferred to the electron system, $\propto c_e/(c_e+c_{ph})$ \cite{vodolazov2017single}. Additionally, smaller $c_{ph}$ reduces thermal noise, caused by thermal fluctuations in electron energy, by decreasing the variance of electron energy fluctuations, in full accordance with statistical thermodynamics \cite{hajdu1977r}. This might play a key role in minimizing the dark count rate in SNSPDs \cite{semenov2020local}, improving their ability to detect extremely low and rare events like dark matter particles \cite{chiles2022new}, and enhancing timing accuracy (timing jitter) critical in laser ranging applications.

The performance of Hot Electron Bolometers (HEBs)\cite{Baselmans2006DirDet}, employed in terahertz spectroscopy, also benefits from these adjustments in thermal properties. The responsivity of HEB-mixers is a key metric in their numeric calibration and is directly affected by the $c_e/c_{ph}$ ratio. Any inaccuracies in defining this ratio eventually lead to calibration errors. For example, in atomic oxygen spectroscopy with HEB \cite{richter20154}, which plays an important role in the chemistry and energy balance of the earth's atmosphere, any calibration errors can skew the forecast for climate change.

Such devices for hybrid electronics as thermal switches (nanocryotrons), which rely on superconducting nanowires \cite{mccaughan2014superconducting, baghdadi2020multilayered} see variations in the heating/switching device operation due to changes in thermal properties induced by confinement. For example, slowed-down electron energy relaxation leads to smaller dissipated heat (heating currents) and slower switching dynamics.

A way to control confinement is the reduction of the device dimensions as in films, wires, or grains. For instance, the discussed performance enhancements are readily achievable in SNSPDs based on amorphous superconductors like MoSi and WSi by reducing the film thickness. In polycrystalline materials such as NbN and NbTiN, reduction in the grain size by adjusting the sputtering conditions (e.g. ambient substrate temperature and gas flow \cite{dane2017bias}) can similarly improve device metrics.

\section{Conclusion}
In this study, we developed and validated a computational model of phonon-mediated heat transport under confinement in nanostructures at low temperatures. Our findings reveal that confinement significantly reduces phonon heat capacities and slows down electron cooling to the substrate, which aligns with the granular morphology of the studied NbTiN nanostructures.

These insights pave the way for future advances in the design and performance of cryogenic nanodevices through tailored microstructure engineering. Notably, the increased infrared sensitivity of superconducting single-photon detectors and improved accuracy in thermal management in Hot Electron Bolometers would significantly impact quantum computing and atmospheric science.

\section*{Acknowledgment}
The authors greatly acknowledge V. Zwiller for his help in the device preparation, A. Sergeev for his essential insights regarding the impact of the magnetic field on electron-phonon interaction, as well as A. I. Bezuglyj for the fruitful discussion on phonon confinement.
M.S. acknowledges funding support from the National Research Foundation, Singapore and A*STAR under the Quantum Engineering Programme (QEP-P1).
A.Z. gratefully acknowledges funding from the European Union through Horizon Europe ERC Grant number: 101043968 ``Multimech'', and from US Army Research Office through contract nr. W911NF-22-2-0256.

\appendix 

\section{Confinement of phonon modes}
\label{app: tau_es}

The confinement model, described in the main text, uses Debye DOS and integration over an angle-dependent wavevector (see e.g., Eq.~(\ref{eq:c_film})). Alternatively, one can introduce the confinement effect on the DOS using the approach of \cite{yu2022omega}. 

The total number of states is found as the ratio of the volume in \textit{k}-space, $\delta V_k= 2 \int_0^{2\pi} \text{d}\phi \int_0^{\pi/2} \text{d}\Theta \,\sin\Theta \, k^2 \text{d}k$ (performing the angular integrals gives $4\pi$), to the volume occupied by a single state, $(2\pi)^3/V$, that is $N=V /(2\pi^2) k^2 \text{d}k$. And the Debye density of states (per unit volume) is simply $g(k)=\text{d}N/\text{d}k=k^2/2\pi^2$. Under confinement imposed by film planes, the minimum allowed wavevector depends on the polar angle as $k_{min}=2\pi \cos\Theta/d$, which modifies the volume in \textit{k}-space as follows $\delta V_k^* = 2 \int_0^{2\pi} \text{d}\phi \int_{\arccos(kd/2\pi)}^{\pi/2} \text{d}\Theta \,\sin\Theta \, k^2 \text{d}k = 2dk^3 \text{d}k$. The modified DOS (per unit volume) is then $g^*(k)=k^3 d/4\pi^3$ \cite{yu2022omega}. 

One then can express phonon heat capacity (per mode) as a sum of capacities with the modified DOS within a sphere of radius $k^*=2\pi /d$ and with the Debye DOS within a spherical shell extending from $k^*$ to the Debye wavevector $k_D$:
\begin{multline}\label{eq:c_film_alter}
    c_{ph}^{\langle2D\rangle}(T, d) = \frac{d}{4\pi^3} k_B \left(\frac{k_B T}{\hbar u}\right)^4 \int_{0}^{x^*} \, \frac{x^5e^x}{(e^x-1)^2} \, \text{d}x \\
        + \frac{1}{2\pi^2} k_B \left(\frac{k_B T}{\hbar u}\right)^3 \int_{x^*}^{x_D} \, \frac{x^4e^x}{(e^x-1)^2} \, \text{d}x,
\end{multline}
where $x^*=2 \pi \hbar u  / d k_B T$. In the limit of strong confinement, $x^* \to x_D$, the second term diminishes, and  $c_{ph} = 120 \zeta(5) \frac{d}{4\pi^3} k_B \left(\frac{k_B T}{\hbar u}\right)^4 \propto T^4 d$ with the Riemann zeta function $\zeta$.

When deriving phonon escape time under confinement imposed by thin film, the heat flow integrals are modified as follows:
\begin{multline}\label{eq:I_2D}
    \mathcal{I}^{\langle2D\rangle}(T)= \int_0^{\pi/2} \alpha(\Theta) \sin\Theta\cos\Theta \\
    \times \int_{\omega^*(\Theta)}^{\omega_D} \frac{g(\omega) \, \hbar\omega}{e^{\hbar \omega/k_B T}-1} \,  \text{d}\omega \, \text{d}\Theta,
\end{multline}
where, $\omega^*(\Theta)=u \, 2\pi \cos \Theta/d$. For a given direction of phonon propagation defined by the angle $\Theta$, only phonons with wavevectors $k>2\pi \cos \Theta/d$ (wavelengths $<d/ \cos \Theta$) are present in the film and may carry the energy out of the film; they have the Debye DOS $g(\omega)$. The derivative of the heat flow integral over temperature is given by:
\begin{multline}\label{eq:dI_cdT}
    \frac{\text{d} \left[ \mathcal{I}^{\langle2D\rangle}(T)\right]}{\text{d}T} = \frac{1}{2\pi^2} k_B \left(\frac{k_B T}{\hbar u}\right)^3 \\
  \times  \int_0^{\pi/2} \alpha(\Theta) \sin\Theta\cos\Theta  
    \int_{x^*(\Theta)}^{x_D} \frac{x^4e^x}{(e^x-1)^2} \, \text{d}x \, \text{d}\Theta \\
    = \langle\alpha\rangle c_{ph}^{\langle3D\rangle}(T)  \biggl[ 1 - \mu(T, d) \biggr],
\end{multline}
where,  $x^*(\Theta)$ and $\mu(T, d)$ were defined in the main text. The integral in Eq.~(\ref{eq:dI_cdT}) are expressed as a difference $ \int_{x^*(\Theta)}^{x_D} \text{d}x= \left[ \int_{0}^{x_D} \text{d}x - \int_{0}^{x^*(\Theta)}\text{d}x \right]$. Finally, the escape time is modified to Eq.~(\ref{eq:tau_es_2d}).

Finally, to describe the experimental data we accounted for all three polarizations (two transversal and one longitudinal) and used the weighted time
\begin{equation}\label{eq:esc_weight}
    \tau_{es} = \left[\dfrac{\sum_i \tau_{es, i}^{-1}\, c_{ph,i}} {\sum_i c_{ph,i}}\right]^{-1}.
\end{equation}
Here the decay rate of each phonon polarisation $\tau_{es,i}^{-1}$ is weighed by its corresponding heat capacity $c_{ph,i}$. This ensures that the contribution of each polarisation to the overall heat transport is proportionate to its ability to store thermal energy. We also note that the acoustic match model\cite{kaplan1979acoustic}, which we used to compute the transmission coefficient $\langle\alpha\rangle$, accounts for mode conversion.

\section{Self-heating normal domain in magnetic field}
\label{app_SHNM}

\begin{figure*}[t!]
	\centerline{\includegraphics[width=1\textwidth]{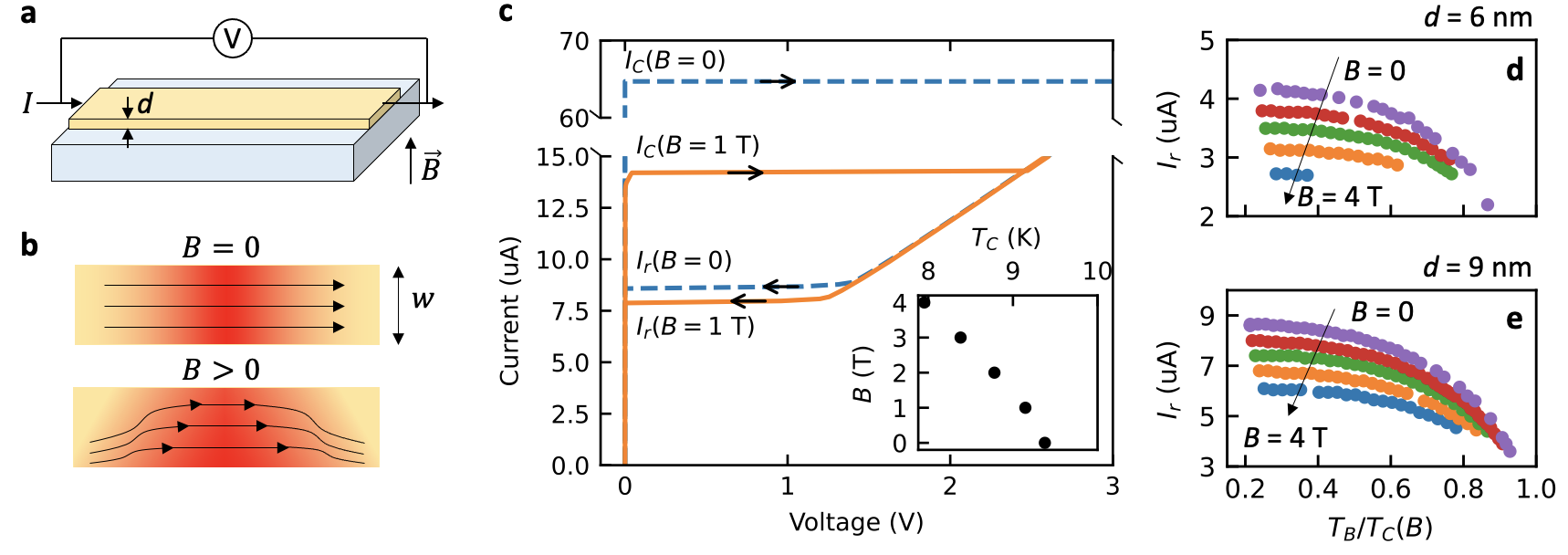}}
	\caption{ (a) Schematic of self-heating domain experiment in external magnetic field $B$. (b) Schematics of the magnetic field effect on the current density (indicated by current lines) and on the domain shape (shown in red). (c) IV curves measured at the bath temperature $T_{bath}=2.4$~K with ($B = 1$~T, solid line) and without ($B = 0$, dashed line) magnetic field for the 9~nm-thick strip. Arrows indicate the sweep direction of the current. $I_C$ and $I_r$ are the critical (switching) and hysteresis (return or retrapping) currents. Inset: Field-dependent transition temperature. (d) and (e) Experimental hysteresis currents at preset magnetic fields ($B$ = 0, 1, 2, 3, and 4~T) vs. bath temperatures ($T_{bath}$) normalized with field-dependent transition temperatures $T_C(B)$ for strips with thicknesses $d=6$ and 9~nm, respectively.}
	\label{fig:Setup}
\end{figure*}

The self-heating normal domain (SHND) technique is an established method\cite{skocpol1974self,dane2022self,yamasaki1979self,sidorova2022phonon} for studying heat transport in superconducting strips. It is done by creating a steady-state normal domain in a superconducting strip heated by a hysteresis current at temperatures above the transition temperature $T_C$. We extended this technique for non-zero magnetic filed to expand the experimental temperature range beyond $T_C$. Below, we provide experimental details and reasoning why the formalism, which we developed in detail in Ref.\cite{sidorova2022phonon} for zero magnetic field is valid for non-zero fields. For mathematical details we refer to Ref.\cite{sidorova2022phonon}.

We studied heat transport in thin NbTiN nanostrips at temperatures below 12~K. Measurements were conducted in physical property measurement system in magnetic fields ranging from 0~T to 4~T applied perpendicular to the substrate plane as sketched in Fig.~\ref{fig:Setup}a. The studied NbTiN strips with thicknesses $d = 6$~nm and 9~nm, widths $w=200$~nm, and lengths $l=100~\mu$m, were fabricated on Si substrates above 270~nm-thick thermally grown SiO$_2$ buffer layers (properties of these NbTiN films were reported in \cite{sidorova2021magnetoconductance}). 

Figure~\ref{fig:Setup}c shows typical hysteretic current-voltage ($IV$) curves of a superconducting nanostrip in zero magnetic field $B=0$ (dashed curve) as well as in a non-zero field of 1~T (solid curve). 
Sweeping the current from the above critical (switching) value $I_C$ downwards makes the nanostrip to return from the normal to the superconducting state through a resistive plateau defined by the hysteresis (return or retrapping) current $I_r$. This plateau corresponds to a self-heating normal domain with a length proportional to the voltage. External magnetic field reduces both currents, $I_C$ and $I_r$, as depicted by the solid curve in Fig.~\ref{fig:Setup}c. It also suppresses the $T_C$ (inset in Fig.~\ref{fig:Setup}c). We note here that the measured $I_r(B=0)$ for the 9~nm-thick strip is consistent with values reported earlier for similar 10~nm-thick NbTiN films \cite{steinhauer2020nbtin} (see also a comment \cite{Ir_substrates}).

Figure~\ref{fig:Setup}d,e, shows $I_r(T_B, B)$ dependencies, which were extracted from $IV$ curves measured at different bath temperatures, $T_{bath}$, and preset magnetic fields. In our experiment, the applied magnetic fields drive the superconducting strip into a mixed vortex state. This state emerges between the upper-critical field $B_{c2}(T=0)>13.9$~T (estimated based in data in \cite{sidorova2021magnetoconductance}) and the field $B_C\approx0.2$~T at which a transition from the vortex-free Meissner to mixed vortex state occurs (estimated according to the Ginzburg–Landau model \cite{maksimova2001critical} and consistent with values reported for similar NbTiN films \cite{jonsson2022current}). Screening currents induced by a non-zero magnetic field sum up with the transport current, resulting in non-uniform net current density across the superconducting part of the strip as illustrated in Fig.~\ref{fig:Setup}b with current lines. Although this may alter the shape of normal domain edges (sketched in Fig.~\ref{fig:Setup}b in red), for large domains, the translational symmetry (space homogeneity) with respect to domain edges dictates that the edges and the mixed vortex state of superconducting parts of the strip do not affect the temperature at the domain center ($T_{e0}$). For sufficiently large domains, the finite temperature at the domain center requires the derivative $\partial^2 T/\partial x^2$ at the center to approach zero. These conditions allow us to apply the SHND formalism, developed \cite{sidorova2022phonon} for $B = 0$, to the non-zero fields.

In the SHND model , the hysteresis current is given by
\begin{equation}\label{eq:current}
I_r = \sqrt{\frac{c_e(T_{e}) w^2 d }{p\:\tau_E(T_{e}) R_\square  T_{e}^{p - 1}}(T_{e}^p - T_B^p)},
\end{equation}
where $T_{e}$, $c_e$, and $R_\square$ are the electron temperature in the domain center, the volumetric heat capacity of electrons, and the sheet resistance of the film, respectively. The steady-state time of electron energy relaxation to the substrate  $\tau_E(T_{e})\approx \tau_{EP}(T_{e}) + \tau_{es} c_e(T_{e})/c_{ph}(T_{ph})$ is given by the advanced two-temperature model\cite{SemenovAnalytical2024}, which is similar to the conventional model\cite{perrin1983response}. Here, $\tau_{es}$ is the phonon escape time, and $\tau_{EP}$ is the electron-phonon (\textit{e-ph}) interaction time, $T_{ph}$, is the phonon temperature in the domain center. 

To validate the use of Eq.~(\ref{eq:current}) for the evaluation of the experimental data at non-zero magnetic field, we first estimate the field impact on the parameters entering this equation. Magnetic field separates the electron states with opposite spin directions by the Zeeman splitting energy $2 \mu_B B$ ($\mu_B$ is the Bohr magneton) that affects the electron density of states $D(E)$ and, consequently, the electron heat capacity $c_e=\int D(E)f_{FD}(E)E dE$ (here $f_{FD}(E)$ is the Fermi-Dirac distribution function). The correction to the electron heat capacity is of the order of $(\mu_B B/E_F)^2/8$ 
where $E_F$ is the Fermi energy. For fields used in our experiment, the splitting energy ($<0.5$~meV) is much smaller than the Fermi energy ($\approx 5$~eV), and therefore the effect of splitting on $D(E)$ and $c_e$ is negligible. Furthermore, it is clear that magnetic field has no direct impact on the number of phonons and, therefore, no effect on $c_{ph}$ and $\tau_{es}$. 

Now let us consider possible field effects on $\tau_{EP}$. In a non-magnetic and non-piezoelectric normal metal, magnetic field can affect the \textit{e-ph} coupling via screening of electromagnetic interaction between electrons and vibrating ions of the crystalline lattice and via induced changes in the deformation potential. The effect through both channels is parameterized with the frequency-dependent conductivity tensor \cite{SPECTOR1967291}. The direct electromagnetic interaction is completely screened in our case, since, for $T_{ph} = 10$~K, frequencies of thermal acoustic phonons $\omega_T/(2\pi) \approx 2\cdot10^{11}$~Hz are much less than the plasma frequency $\omega_p/(2\pi) \approx 4\cdot10^{15}$~Hz and the characteristic frequency $\sigma^{-1}\varepsilon_0  \varepsilon_r/(2\pi)\approx 2\cdot 10^{15}$~Hz ($\sigma$ and $\varepsilon_r$ are the dc conductivity and the dielectric permittivity at infinite frequency, respectively) above which the conductivity becomes frequency dependent. 
On the other hand, the applied magnetic fields are not strong enough to cause cyclotron resonance (Landau quantization). Indeed, in our case the cyclotron frequency $\omega_c = e B/m_e$ ($\omega_c/(2\pi) \approx 1.1\cdot10^{11}$~Hz for $B=4$~T) is much smaller than the reciprocal elastic scattering time $\tau\approx 1$~fs \cite{sidorova2021magnetoconductance}. 
Under these conditions, electrons do not complete at least one entire orbit without being scattered and, consequently,
cannot experience a resonance-energy exchange with thermal phonons. Corrections to the diagonal elements in the conductivity tensor \cite{abrikosov2017fundamentals} are of the order $(\omega_c\tau)^2 \approx 1.5\cdot10^{-6}$ and can be safely neglected.
Finally, since all the discussed material parameters do not depend on the magnetic field, the exponent $p$ in Eq.~(\ref{eq:current}) is likewise field-independent (in a particular film $p$ is determined by the phonon dimensionality and the degree of disorder \cite{bezuglyi1997kinetics}).

Assuming field-independent material parameters, the experimental $I_r(T_B, B)$ data, shown in Fig.~\ref{fig:Setup}(d,e), were fitted with Eq.~(\ref{eq:current}). The temperatures $T_{e}$ and $T_{ph}$ were found for each pair of values $B$ and $T_{bath}$. Other material parameters for our films were taken from Ref.\cite{sidorova2021magnetoconductance}. We used $\tau_E$ and $p$ as the only field-independent fitting parameters. The fit was subject to an additional restriction: $\tau_E$ to be a smooth function (power law) of temperature. The best-fit $\tau_E(T_{e})$ obtained with the least-squares method are shown in Fig.~\ref{fig:exp1}a.

Best-fit values of exponents $p=4.2$ and 3.6 were found for strips with thicknesses $d=6$~nm and 9~nm, respectively. Fitting $\tau_E(T_{e})$ data with a power law  (solid curves in Fig.~\ref{fig:exp1}a) results in dependences $\tau_E \propto T^{-1.9}$ for the 6~nm-thick film and $\tau_\epsilon \propto T^{-1.4}$ for the 9~nm-thick film.

\bibliography{manuscript_return_currents/phonon_heat}

\end{document}